# Revealing travel patterns and city structure with taxi trip data

Xi Liu[a,b], Li Gong [a,b], Yongxi Gong [c], Yu Liu [a,b*]

[a] Institute of Remote Sensing and Geographical Information Systems, Peking University, Beijing 100871, PR China

[b] Beijing Key Lab of Spatial Information Integration and Its Applications, Peking University, Beijing 100871, PR China

[c] Shenzhen Key Laboratory of Urban Planning and Decision Making, Harbin Institute of Technology Shenzhen Graduate School, Shenzhen 518055, PR China

**Abstract**: Delineating travel patterns and city structure has long been a core research topic in transport geography. Different from the physical structure, the city structure beneath the complex travel-flow system shows the inherent connection patterns within the city. On the basis of massive taxi trip data of Shanghai, we built spatially-embedded networks to model the intra-city spatial interactions and introduced network science methods into the issue. The community detection method is applied to reveal sub-regional structures, and several network measures are used to examine the properties of sub-regions. Considering the differences between long- and short-distance trips, we reveal a two-level hierarchical polycentric city structure of Shanghai. Further explorations on sub-network structures demonstrate that urban sub-regions have broader internal spatial interactions, while suburban centers are more influential in local traffic. By incorporating the land use of centers from the travel pattern perspective, we investigate sub-region formation and center–local places interaction patterns. This study provides insights into using emerging data sources to reveal travel patterns and city structures, which could potentially aid in applying urban and transportation policies. The sub-regional structures revealed in this study are more easily interpreted for transportation-related issues than other structures, such as administrative divisions.

**Keywords**: GPS-enabled taxi data; travel pattern; urban structure; spatially-embedded network; community detection

## 1. Introduction

The structures of cities are closely related to the intra-city travel patterns of their residents. The allocation of resources within a city generates travel demands and drives people to travel, while the travel flows inversely indicate the need for modification of current living habitats and construction of new transport facilities. On the basis of approaches such as field surveying, remote sensing, and policy consulting, urban forms are more accessible than travel data. Thus, compared with the impacts of travel behaviors on city structure, more studies focus on the impacts of city structure on travel behaviors, which has drawn the attention of geographers since the 1970s (Yue et al., 2014). Researchers have investigated the influence of urban form from different aspects, such as the mixture of land uses and settlement sizes (Stead and Marshall, 2001; Song et al., 2014), and have tried to evaluate urban policies through travel behaviors of residents (Schwanen et al., 2004; Lowry and Lowry, 2014). Handy (1996) categorized the methodologies of these studies into five categories: simulation studies, aggregated analyses, disaggregated analyses, choice model, and activity-based analyses. Among them, aggregated analyses focus on the aggregated-level characteristics of individual travel behaviors, which are referred to as travel patterns.

Although urban forms can be relatively easily measured, we should nevertheless interpret the cities beyond the spatial distributions of their physical environments and economical resources. The underlying structure of a city, such as which regions have more internal spatial interactions and how the city centers interact with their vicinities, illustrates the method for cities to function as dynamic systems rather than as static artifacts. The ties that connect the discrete physical resources of a city into an integrated system are flows of people and freight, and the flows represent the spatial interaction strengths between places. Studies have been conducted to reveal the underlying city structure via flow systems since the 1960s, when Berry (1966) explored the spatial structure beneath the complex flow systems. Because of the limitations of data sources, analytic tools, and computation capabilities (Yue et al., 2014), these studies had limited development and most studies relating to city structures still focused on urban morphology. Travel behavior studies experienced a low period during the 1990s (Timmermans et al., 2002) for the same reason.

Technological achievements in recent years have once again brought travel patterns and urban structure to the forefront of transport geography research. Big geospatial data, which are collected from sources such as mobile phone records, social media check-ins, and taxi trajectories, provide abundant locations to model movement of people around cities. Compared with traditional travel survey data, big data are more accurate, objective, plentiful, cost-effective, and accessible, and they provide opportunities to better describe people's movements (Lu and Liu, 2012). Big data have also drawn the attention of statistical physicists (Brockmann et al., 2006; Gonzalez et al., 2008; Schneider et al., 2013; Song et al., 2010) and computer scientists (Cho et al., 2011) to provide novel ideas on issues related to human mobility and city structure. The massive travel flows enable us to model an entire city into a spatially-embedded flow network (Batty, 2013). Armed with recently-developed network analytic methods, we are able to explore the sub-regional structures of cities and the patterns of people traveling within those sub-regions. Community detection has been developed to find sub-structures of networks, and it divides a network into sub-networks that have stronger connections within themselves than with others. In terms of flow networks, sub-networks represent sub-regions that have strong internal spatial interactions. Many researchers have applied community detection and other network science methods on a national scale trying to determine whether existing administrative boundaries are still reasonable (Ratti et al., 2010; Thiemann et al., 2010; Liu et al., 2014), exploring the relations between commuting properties and socio-demographic variables (De Montis et al., 2007), comparing human spatial interactions within different countries to find common patterns (Sobolevsky et al., 2013), and providing suggestions for regional partitions (De Montis et al., 2013).

While large volumes of movements can be extracted from big geospatial data, most city-level studies still treat the origination and the end of a trip as two unrelated activities attempting to, for example, depict the city structure from the land use perspective (Guo et al., 2012; Liu et al., 2012b; Toole et al., 2012; Reades et al., 2009). With regard to studies that systematically view intra-city flows, Tanahashi et al. (2012) applied graph-partitioning methods to the human mobility network extracted from phone records in New York City, focusing on human travel patterns between the partitioned sub-regions instead of revealing regional structures. Roth et al. (2011) utilized individual travel information of the London Underground to explore the polycentric city structure, but the subway travels were constrained by linear routes, and thus, the results are unable to fully reflect the urban dynamics.

We apply taxi trip data, which have been widely used to investigate city structure since the 1970s (Goddard, 1970), to implement our method. The recent availability of large volumes of taxi Global Positioning System (GPS) trajectories has strongly promoted related studies, including transportation analysis (Fang et al., 2012; Gao et al., 2013; Li et al., 2011), urban planning (Veloso et al., 2011; Zheng et al., 2011), land use analysis (Kang et al., 2013; Liu et al., 2012b; Qi et al., 2011; Yuan et al., 2012), human mobility patterns (Liang et al., 2012; Liu et al., 2012a), and spatial model calibration (Yue et al., 2012). Although taxi trajectories are unable to reflect continuous displacements of specific people, which are crucial for the time geography framework, they describe collective human mobility patterns of intra-city travels with accurate positions and time stamps. With information of when and where a customer is picked up or dropped off by a taxi, meaningful trips corresponding to displacements between people's consecutive activities are easy to extract, which is a difficult and time-consuming effort to obtain from other forms of data. Precise spatiotemporal properties of massive intra-city trips generated by taxi trajectories lay a solid foundation to completely reflect the structure of the city.

In this study, we introduce complex network science methods to analyze GPS-enabled taxi data collected in Shanghai, China and explore the structure of intra-city flows. To comprehensively reveal the city structure from different levels, we pay much attention to patterns of trips for different length. The contribution of this work is twofold. First, our knowledge of the city structure mostly comes from urban designs and plans, which are partially arbitrary. However, this research provides an objective bottom-up view to depict the structure with residents' travel flows. It extends the exploration of complex city flow networks with new data and methods and demonstrates the cities' function in reality. Second, with taxi data acting as an observatory to intra-urban flows, the city structure revealed in this study has strong connections with transportation applications. Further explorations of different flow structures in urban and suburban sub-regions provide new insights for traffic optimization and city management. The rest of the paper is organized as follows: Section 2 describes the study area and the preparation of data; Section 3 introduces the community detection method, reveals the hierarchical polycentric city structure, and explores the land use of the centers; and discussion and conclusions, along with suggestions for policy-making, are stated in Section 4.

# 2. Study area and data preparation

## 2.1 Shanghai districts

As the economic center of China, Shanghai (Fig. 1) is an international-level metropolis, with a land area of more than 6,000 km$^2$. It comprises 16 districts and 1 county since 2012. The eight districts in Puxi, along with the Lujiazui area in Pudong, are the core urban areas of Shanghai (shown in the rectangle in Fig. 1). Pudong District is separated from the western part of Shanghai, the older urban area, by the Huangpu River and has experienced rapid development in recent decades. Shanghai has two airports, viz., the Shanghai Pudong International Airport and the Shanghai Hongqiao International Airport. Shanghai Railway Station and Shanghai South Railway Station were two major railway hubs in 2009. The study area covers all the Shanghai districts except for Chongming County.

The public transit system, including buses, subways, taxis, and ferries, serves a large part of intra-city travels in China. In 2010, the public transport system catered to 34% travels in Shanghai and 47% travels in the core urban area, with taxis producing 19.3% of the intra-urban trips by public transportation[1]. Thus, taxis form an important supplement to buses and metros in Shanghai. Traveling by taxi offers flexible routes and is more time-efficient than other modes of transportation. In China, taxi trajectories provide a reasonable data source for urban studies, given their capacity to capture a large proportion of urban passenger flows.

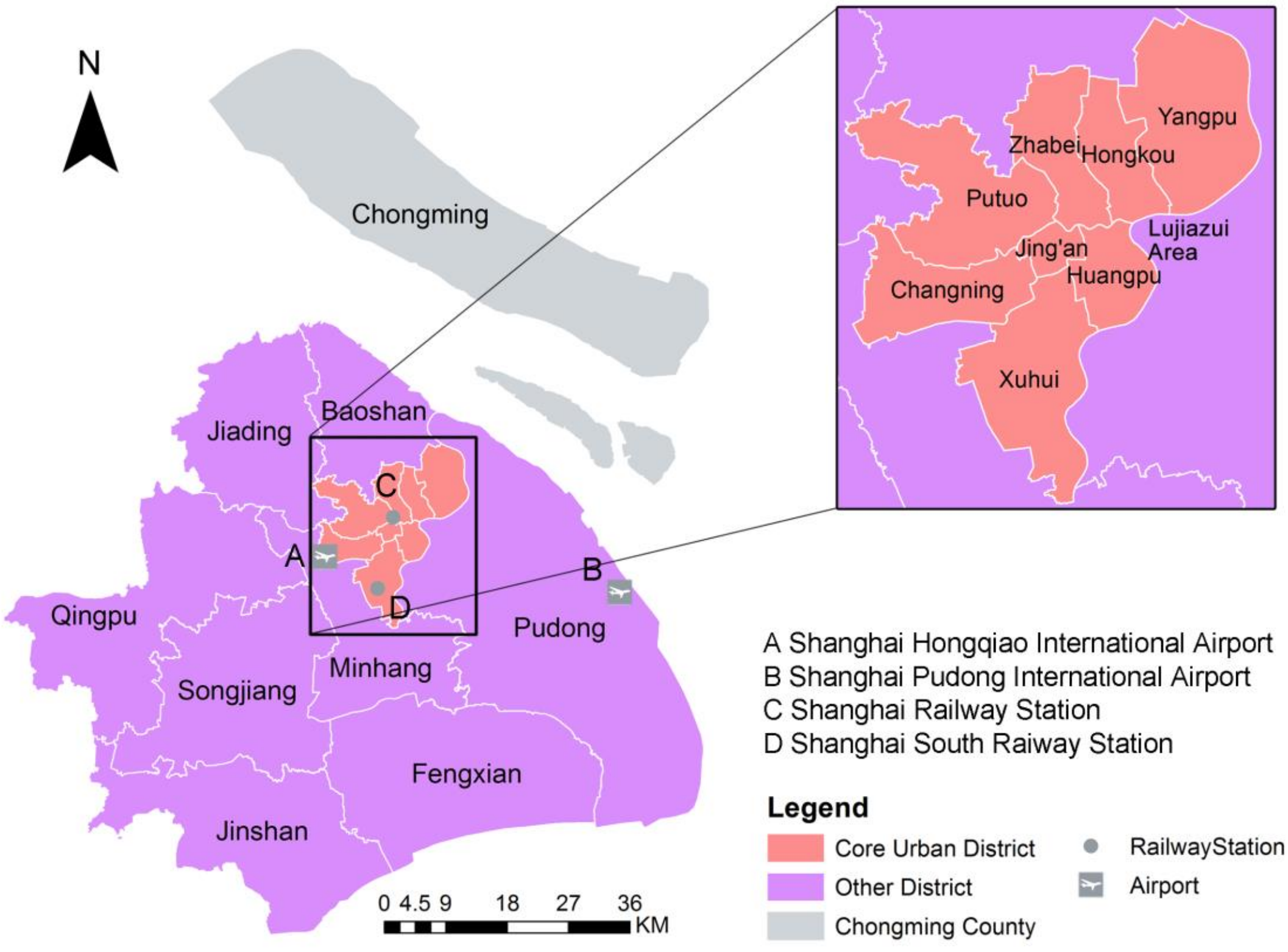

1 http://www.shanghai.gov.cn/shanghai/node2314/node25307/node25455/node25459/u21ai605264.html, in Chinese.

Fig. 1. Shanghai, the study area. The red districts, along with the Lujiazui Area of Pudong, form the core urban area of Shanghai. The purple districts and Chongming (not included in the study area) are the suburban and rural districts.

## 2.2 Taxi trip data and data preparation

The source data set contains GPS trajectories of more than 6,600 taxis in Shanghai. The trajectories cover the entire city, while the central area concentrates relatively higher trip volumes. Detailed descriptions of the data set can be found in Liu et al. (2012a). Taxi trips on weekdays are more appropriate for revealing urban structure because people tend to travel more regularly on weekdays, and data recorded on Friday was excluded as people take more trips for entertainment on Friday night. Therefore, we used a data set that covers four consecutive days from Monday (June 1, 2009) to Thursday (June 4, 2009). Given that we focused on the interactions between places, we simplified taxi trajectories into vectors consisting of origins and destinations. We extracted 860,905 taxi trips from the data set after removing the trips that were either erroneous or extended beyond the administrative boundaries. Each trip record contains information of taxi ID, pick-up time, pick-up point coordinates, drop-off time, drop-off point coordinates, and trip length. This research uses actual trajectory distances instead of Euclidean distances between origin and destination as the length of the trip because people are sensitive to the relatively-expensive price of taxi trips, which correlates with the actual trajectory distance.

# 3. Revealing the two-Level hierarchical polycentric city structure

## 3.1 Network construction and community detection

People's movement flows within a city connect discrete places into an integrated system. Although we are aware of the exact positions a customer is picked up or dropped off by a taxi, the exact place or building the customer comes from or goes to is unknown. Given that a small spatial unit usually has the same land use, we can aggregate trips to obtain spatial interactions among these small regions. The small units could be traffic analysis zones (TAZs), grids, or parcels segmented by major roads. By treating the units as nodes and the movement flows as edges, we are able to construct spatially-embedded networks and apply complex network methods to further study their properties and structures.

To build a network from taxi trip set $\boldsymbol{T}$, the study area was divided into small units with each unit $C_i$ corresponding to a node $V_i$. Trips between two nodes indicate the existence of an edge, or a linkage, between them. If there were $n$ trips originating from $C_i$ and ending in $C_j$, the weight of edge $E_{ij}$ from $V_i$ to $V_j$ was $n$. Thus, a weighted and directed network $N$ was formed. We illustrate the steps in Figs. 2 (a), (b), and (c). In this study, we partitioned the study area into 1 km $\times$ 1 km cells as the nodes of the networks because of the lack of TAZ data. This scale is determined on the basis of relevant studies (Liu et al., 2012b) that suggest that the cell size is detailed enough to depict the urban structure. Additionally, the cells have a size similar to TAZs, acting as appropriate substitutions to represent relatively uniform socio-economic characteristics.

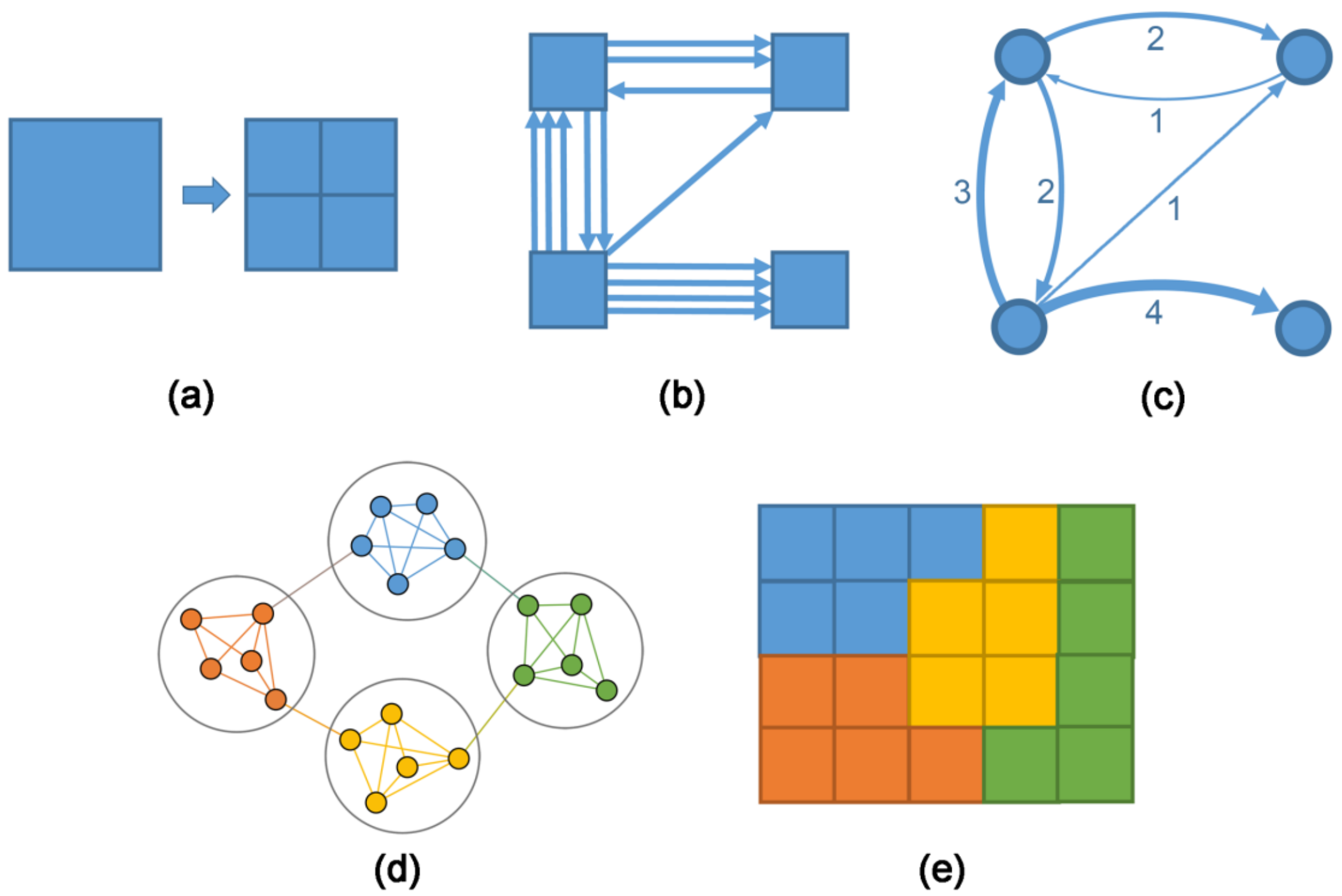

Fig. 2. To construct a network from taxi trips, the study area was divided into small regions (a) with each small region corresponding to a node in the network. A directed edge or linkage existed between two nodes if there were taxi trips from one node to the other. The weight of an edge equals the taxi trips it represents (b, c), and the properties of the network thus reflected the structure of the city and the spatial interactions among the small regions. (d) provides an illustration of the communities detected from a network, which was divided into four parts (depicted by four circles) in which the sub-networks had relatively dense connections. The community detection result corresponds to closely connected sub-regions (e).

In a network, some nodes indicate much stronger connections among them than with other nodes. By dividing the taxi trip network into densely-connected sub-networks, we were able to divide the city into intensely-interactive sub-regions. In network science, community detection methods are able to partition an entire network into tightly connected sub-networks (Fig. 2 (d)), called communities, and reveal the network's clustering characteristics (Girvan and Newman, 2002). Community detection can be implemented using many algorithms, among which the Infomap algorithm is able to handle the weighted and directed networks and performs stably and quickly (Fortunato, 2010). It applies a two-level description of a random walk on a network and aims to minimize the expected description length, or "code length," of the random walk. The optimized two-level description corresponds to the community structure of the network. For detailed information of the Infomap algorithm, please refer to Rosvall and Bergstrom (2008). We adopted the Infomap toolkit provided in igraph R package (Csárdi and Nepusz 2006) for this study.

## 3.2 Two-level hierarchical city structure

We first constructed a network $N_{all}$ comprising all of the taxi trips in the data set. The community detection result of $N_{all}$ (Fig. 3) mainly consists of large, spatially continuous regions indicating the influence of the distance decay effect on spatial interactions. Some single cells in suburban areas also belong to the large urban communities because of some infrequent long-distance travels between the cell and the urban area. Small communities primarily exist in suburban and rural areas because of low taxi trip volume. In Fig. 3, communities with less than 10 cells are not displayed.

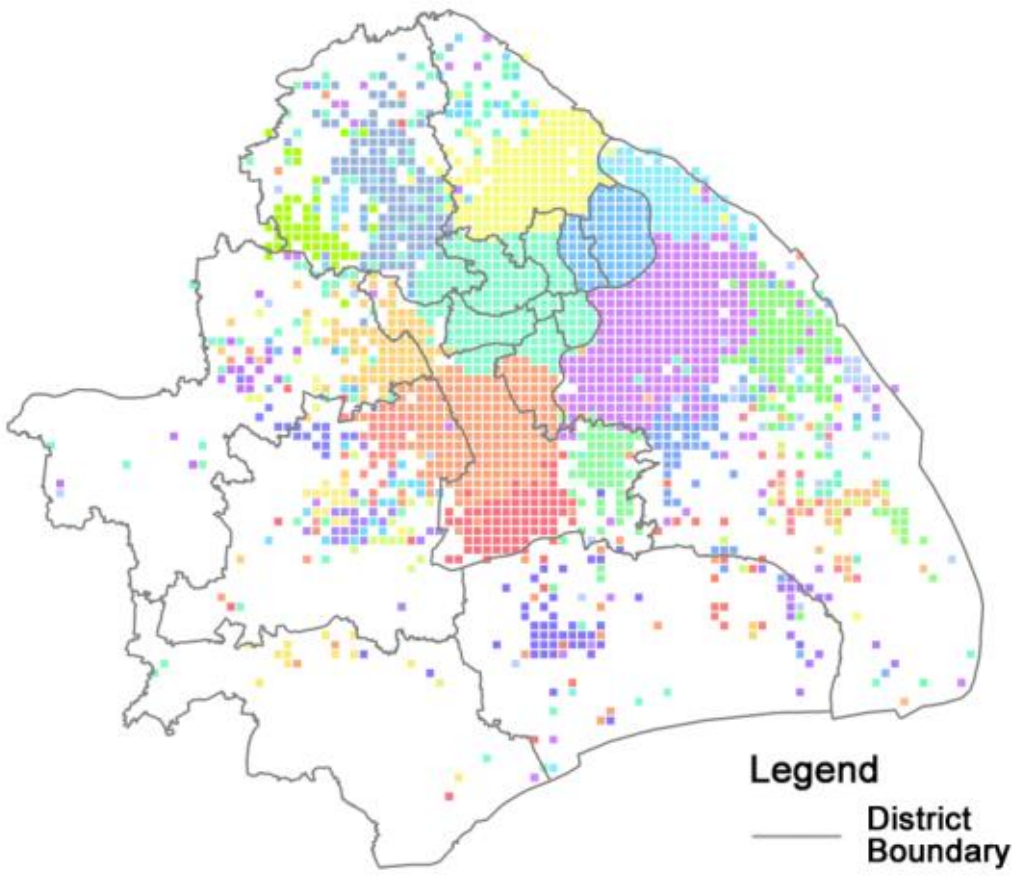

Fig. 3. Community detection result of the network constructed by all taxi trips. Cells in the same color are of the same community.

As shown in Fig. 3, communities in urban areas are large and detailed structures might be concealed. To detect detailed structural information, we paid more attention to short trips for the following three aspects. First, short trips link local places and may help find smaller zones. Second, short and long intra-city travels often have different purposes and travel patterns, especially in a large city such as Shanghai. Third, a majority of taxi trips are short (half of the trips are shorter than 4.72 km and three quarters of the trips are shorter than 8.32 km). They represent the spatial interactions between close places and can, thus, depict the city structure on a more detailed scale. The different patterns between short- and long-distance trips could be confirmed by some simple analyses. For example, splitting all trips into two sub-sets by the median length (4.72 km) and naming them the short-trip set $\boldsymbol{T_s}$ and the long-trip set $\boldsymbol{T_l}$ results in their volumes displaying different temporal distributions (Fig. 4). Short trips reach a higher volume peak at approximately 2 p.m., while long trips have larger volume at night. The spatial distributions of the sum of pick-up points and drop-off points in $\boldsymbol{T_s}$ and $\boldsymbol{T_l}$ (Fig. 5) are also different. More trips in $\boldsymbol{T_l}$ originate or end in regions containing railway stations and airports.

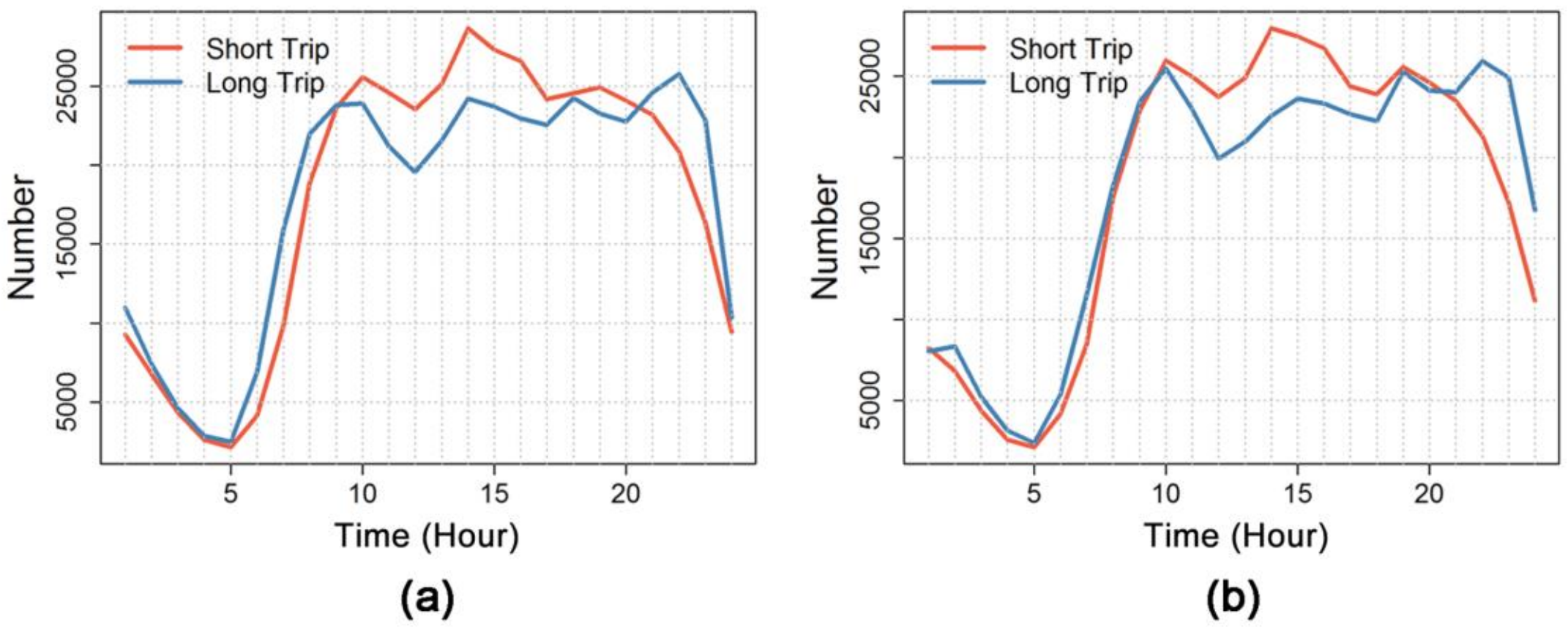

Fig. 4. Temporal pick-up point volumes (a) and temporal drop-off point volumes (b) of short- and long-distance trips. The curves of short and long trips peak at different times.

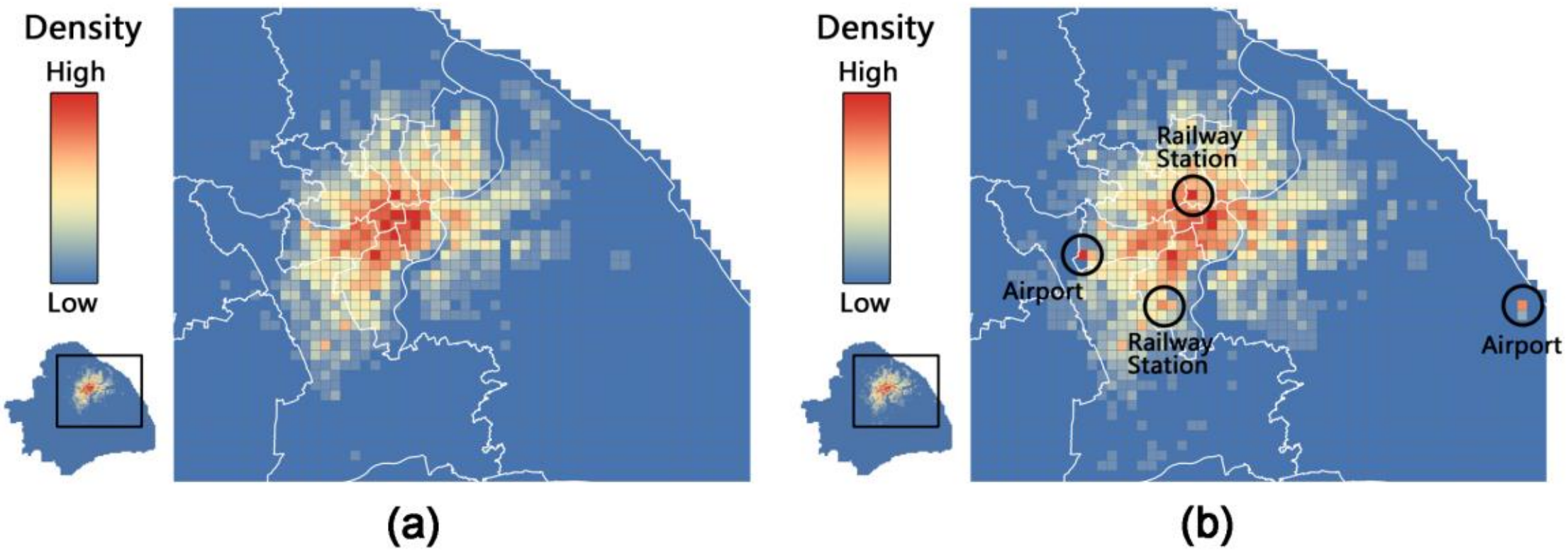

Fig. 5. Spatial distributions of the sum of pick-up points and drop-off points. Hot spots of short trips (a) concentrate in urban areas, whereas hot spots of long trips (b) also include airports and railway stations.

Network statistics could also distinguish the patterns of short and long trips. We constructed networks $N_s$ and $N_l$ from trips in $\boldsymbol{T_s}$ and $\boldsymbol{T_l}$. The two networks contain approximately the same number of trips; however, $N_l$ includes more nodes and less average edge weight. Most of the long-trip network edges (78%) are aggregated by three trips or less, which are, in general, ad hoc and imply greater randomness than short trips. Reasonably, long trips connect places that are far from each other, leading to large sub-regions in the community detection result. Short trips are more spatially stable, suggesting that a more meaningful structure can be identified from $N_s$.

Table 1 Basic statistics of networks constructed from short and long trips

| | Trip number | Nodes | Edges | Average edge weight | Proportions of edges with weight equal or less than 3 |
|---|---|---|---|---|---|
| Short | 430,453 | 2,066 | 29,277 | 14.70 | 0.47 |
| Long | 430,452 | 2,816 | 137,490 | 3.13 | 0.78 |

To reveal the underlying city structure, we first detected clustered sub-networks from the network constructed by short-distance trips, finding a basic structure, and then gradually added long-distance trips into the network, exploring how these long distance interactions influence the city structure. We use $\boldsymbol{T_i}$ to represent the trip data set containing all trips shorter than $i$ km, and $N_i$ is the network constructed from $\boldsymbol{T_i}$. We started by applying the community detection method to $N_2$ because traveling by taxi in Shanghai is rarely less than 1 km (1.6% of total taxi trips). Then, we detected sub-regional structures on $N_3$, $N_4$, $N_5$, ..., and finally $N_{all}$.

The gradually changing results of sub-regional city structures are rather interesting (Fig. 6). They clearly depict a two-level hierarchical structure. The result of $N_2$ consists of small, spatially consecutive regions. When adding longer trips into the network, the small regions start to grow larger and merge with adjacent sub-regions, as illustrated by the results of $N_3$ and $N_4$. However, the results of $N_4$, $N_5$ and $N_6$ are very similar, with few sub-regions expanding or merging. This steady state shows the relatively regular spatial structure of short trips and may result from the concentrations of local travels. The steady state, however, begins to change when longer trips are added, but the sub-regions in the community detection results of $N_7$, $N_8$, ..., and $N_{all}$ are merged by the steady state communities almost without boundary changes, especially in urban areas where trips are abundant. This pattern implies that short trips dominate local spatial interactions, whereas long trips play the role of connecting these local clusters instead of following the same pattern of short trips and maintaining the augmenting process of sub-regions. The low trip volumes in some suburban or rural areas make it difficult to maintain local interaction structures when longer trips are added, leading to some small changes of sub-region boundaries. Note that suburban and rural communities containing less than 10 cells because of low trip volumes have been omitted in the visualizations.

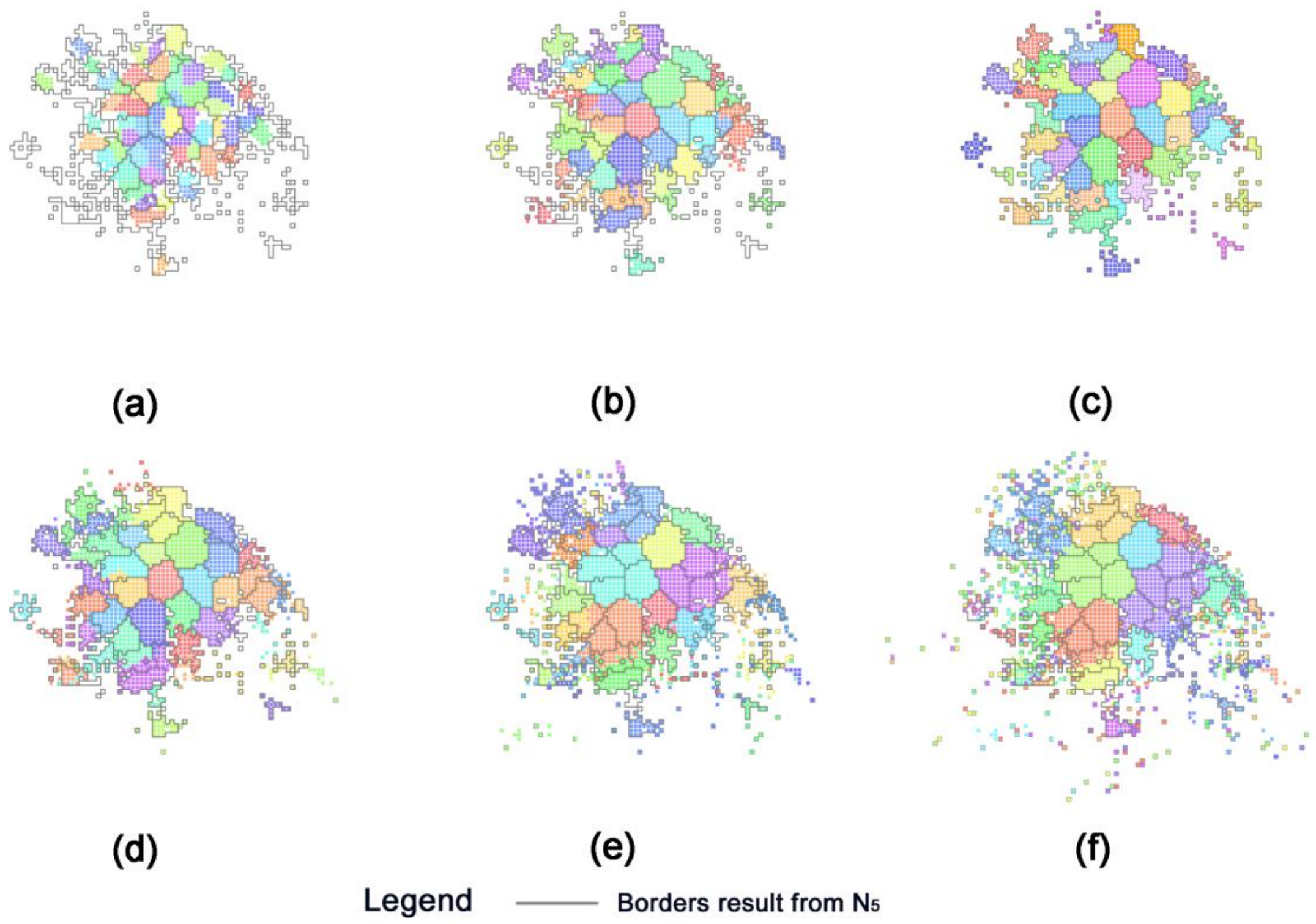

Fig. 6. The community detection results of $N_2$ (a), $N_4$ (b), $N_5$ (c), $N_6$ (d), $N_{12}$ (e) and $N_{all}$ (f). The boundaries of $N_5$ are also depicted in other figures.

Finally, a two-level hierarchical city structure was revealed from the aspect of intra-city interactions. The first-level sub-regions show a steady structure dominated by short-distance spatial interactions, which strongly relate to the travel demand of local residents and their periodic mobility routines. We named these basic sub-regions Level One Zones (L1Z). The community detection result of $N_5$ is used to represent the L1Zs because the result of $N_4$ still grows a little compared with the result of $N_5$, while a few rural L1Zs start to merge in the result of $N_6$. The second-level sub-regions illustrate the interaction patterns reflected by all the taxi trips, and we named these regions Level Two Zones (L2Z). We use the community detection result of $N_{all}$ to illustrate the L2Zs. The hierarchical sub-regional structure provides insights into how the city could be properly divided into closely related sub-regions on the basis of intra-city travels.

The most common way people interpret the sub-regional structure of a city is to refer to its administrative divisions (Ratti et al., 2010; De Montis et al., 2013), but with the exception of the boundary of the Huangpu River between Pudong and the core urban districts in Puxi, most L1Z (Fig. 7 (a)) and L2Z (Fig.4) boundaries are not consistent with district-level administrative boundaries. However, the mismatch between boundaries is reasonable. First, the district boundaries have been arbitrarily determined in a top-down fashion, aiming at facilitating administration rather than considering human mobility. Second, advanced transport facilities, low travel costs, and diverse travel needs result in stronger dynamics in urban areas, which undermine the influence of administrative boundaries on people's travel behaviors. In the rural and suburban districts, where human dynamics are less intensified than the core urban districts, most of the L1Z boundaries are consistent with town/subdistrict[2] boundaries (Fig. 7 (b)), a sub-level administrative division of district in Shanghai. This is a very interesting finding, suggesting that administrative hierarchy shapes intra-urban movements, especially in less developed areas. Such a phenomenon may be traced back to the urban planning a few decades ago, when administrative units played an important role in China's management system. It also demonstrates the reasonableness of the revealed structure.

Therefore, by emphasizing the different semantics behind travels of different lengths, we unveiled the hierarchical structure buried in taxi trip data. The matching boundaries of L1Zs and L2Zs also inversely indicate the different travel patterns considering travel length and their influence on city structure. Compared with administrative boundaries, the hierarchical structure extracted from massive intra-city travel data provides a bottom-up view of our living habitat, throwing light on transportation-related issues. To aid in urban and transportation planning, the underlying mechanism of the structure formation should be explored.

---

[2] Jiedao in Chinese, detailed information can be found in http://en.wikipedia.org/wiki/Subdistricts_of_the_People's_Republic_of_China

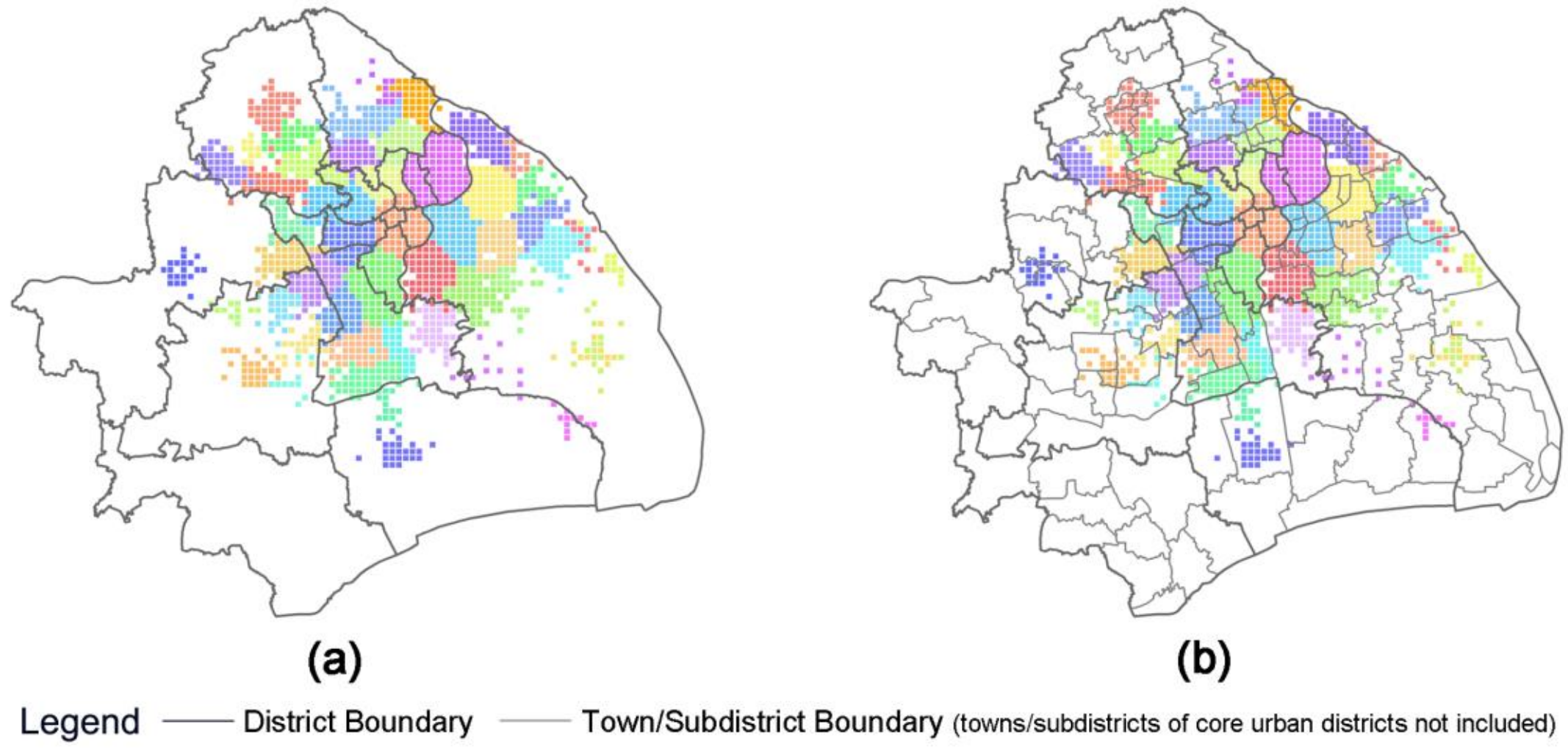

Fig. 7. Comparison of L1Zs and administrative boundaries of Shanghai at the district level (a) and the town level (b). Towns and subdistricts are the sub-level administrative divisions of the districts in Shanghai. Note that the eight core-urban districts are still represented at the district level in (b). The boundaries of many suburban and rural L1Zs are consistent with some town/subdistrict boundaries.

## 3.3 Properties and centers of sub-regions

To explore the two-level sub-regional structure formation, we investigated the sub-networks corresponding to those sub-regions. The sub-networks also depict people's travel patterns in sub-regions. For the first-level sub-regions, 15 L1Zs containing more than 1,000 taxi trips each were selected for further study (Fig. 8 (a)) because too few taxi trips may not guarantee steady structures. These 15 L1Zs cover the core urban area of Shanghai and some suburban sections where traffic problems, such as congestion, are prone to exist. The 15 L1Zs are numbered in descending order according to their internal trip amounts. L1Z 1 lies in the center of Shanghai and contains the largest number of taxi trips. L1Zs from 1 to 7 lie in the urban area, while L1Zs from 8 to 15 lie in suburbs. Urban L1Zs generally contain more trips than suburban trips (Fig. 9 (a)). Moreover, the central urban, urban, and suburban L1Zs lie in a relatively concentric order (the separation of the eastern part and the central area by the Huangpu river caused some modification), which is similar to the Burgess concentric model (Burgess, 1925). For the second-level sub-regions, four L2Zs (Fig. 8 (b)) merged by more than one L1Zs were selected. They are L2Z I, consisting of L1Zs 1, 3, 6, and 7; L2Z II, containing L1Zs 5, 8, 9, and 11; L2Z III, covering L1Zs 4, 10, and 14; and L2Z IV, including L1Zs 12 and 15. Note that L1Zs 2 and 13 remain unmerged at the second level. This may be attributed to the well-developed facilities for daily needs in L1Z 2 and the function as a free trade zone of L1Z 13. To construct sub-networks corresponding to the 15 L1Zs, we used all the internal taxi trips in each L1Z. For the L2Zs, we used internal trips of each L2Z with travel length longer than 6 km, which contribute mostly for merging the L1Zs to form L2Zs.

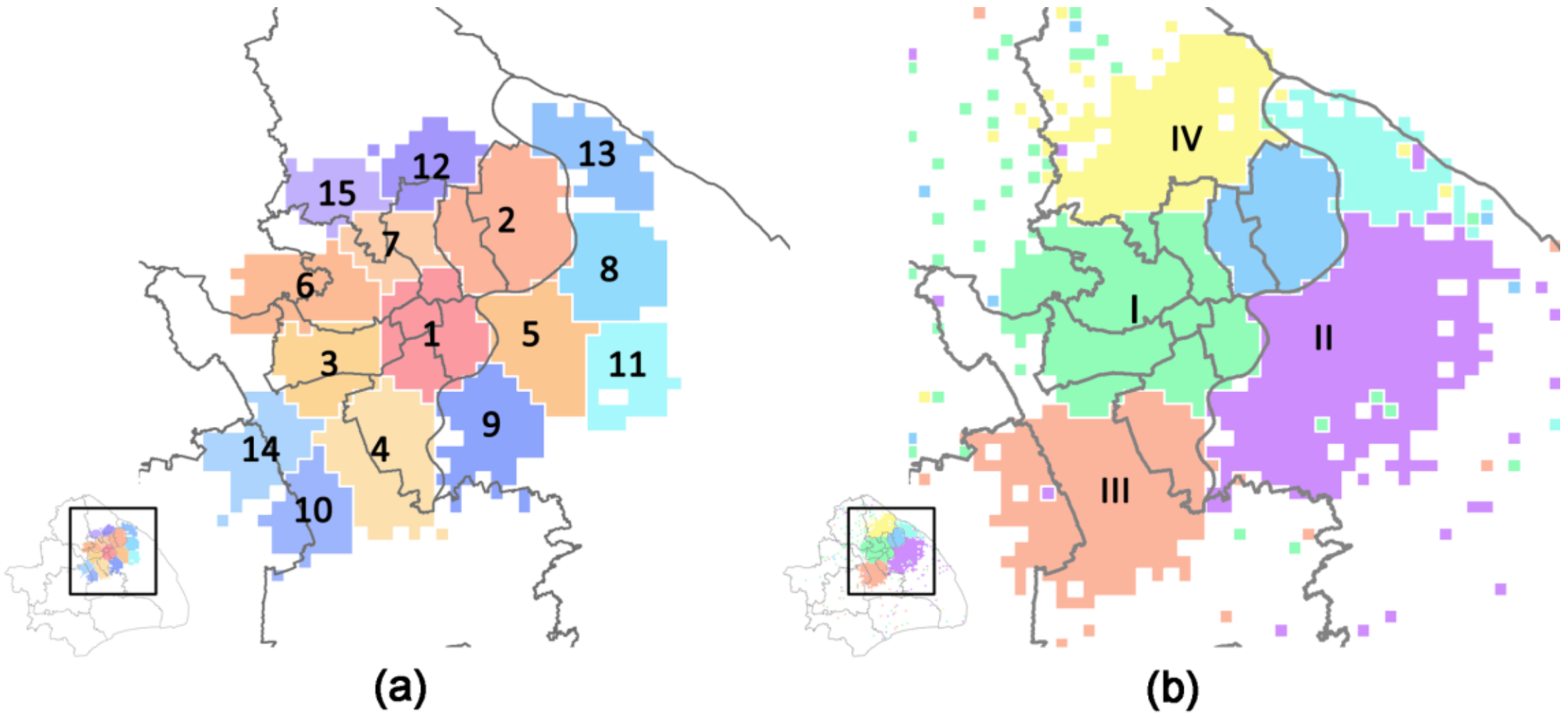

Fig. 8. (a) The spatial distribution of 15 Level One Zones (L1Zs) with more than 1,000 internal trips each. L1Zs from 1 to 7 cover the major urban area of Shanghai, and L1Zs from 9 to 15 lie in the suburbs. (b) The spatial distribution of selected Level Two Zones (L2Zs).

We first dug into sub-networks of the L1Zs with properties including graph density and node strength, which can depict the general spatial interaction patterns among cells in each L1Z and the local travel patterns of people. In network science, graph density (Wasserman, 1994) measures density of intra-node interaction in a network. It is defined as $D = \frac{E}{N(N-1)}$, where $E$ is the number of edges and $N$ is the number of nodes in a graph. The node strength is denoted by its total traffic volume, including both inflows and outflows in directed networks. All 15 L1Z sub-networks have relatively high graph density (Fig, 9 (b)), indicating that the cells in the same L1Z are densely connected. The complementary cumulative distribution function (CCDF) curves of node strength in each L1Z (Fig. 9 (c)) decrease more quickly with small values and more slowly with large values, illustrating that large proportions of nodes have relatively small traffic volumes, whereas a few nodes play roles as centers attracting or generating a large proportion of the total traffic volume. Note that to compare the curves of different L1Zs, we normalized the strengths by dividing them with the maximum strength in each L1Z. To validate this characteristic, we calculated the proportions of the internal trips the L1Z centers take. For simplicity, the node with the largest strength in each L1Z was defined as the local center. As shown in Fig. 9 (d), the centers are associated with significantly large proportions of local trips. By concentrating large amounts of travel, the centers may play important roles in forming the observed spatial structure.

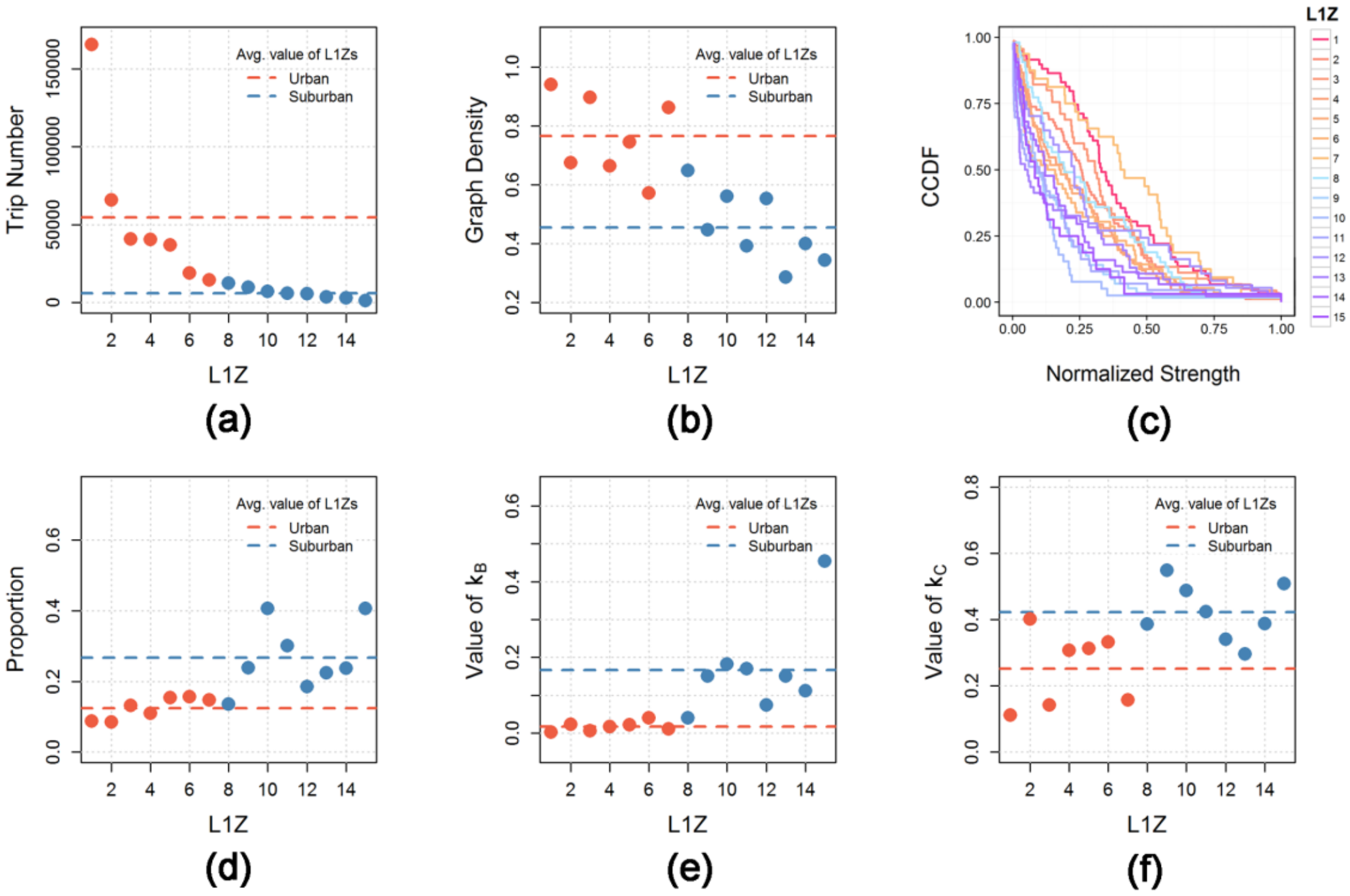

Fig. 9 (a) Total trip numbers in each L1Z. (b) Graph density of each L1Z sub-network. (c) Complementary cumulative distribution function (CCDF) curves of normalized strength in each L1Z. (d) The proportion of total internal traffic amount the centers take in each BTZ. (e) The values of $k_B$ in each L1Z. (f) The values of $k_C$ in each L1Z. The comparisons between average values of urban and suburban L1Zs demonstrate that urban L1Zs have broader interactions, while suburban L1Z centers are more influential in local traffic.

High node strength does not guarantee that these centers are also centers in the aspect of sub-network structure. Thus, we applied two additional node centrality properties: betweenness centrality, which measures the importance of a node in organizing flows in the network, and closeness centrality, which measures the average shortest distance between the node and all the other nodes in the network. We applied the normalized properties in this study. Normalized betweenness centrality is defined as $C^B(i) = \frac{1}{(N-1)(N-2)} \sum_{s \neq t} \frac{\sigma_{st}(i)}{\sigma_{st}}$, where $N$ is the total number of nodes, $\sigma_{st}(i)$ is the number of shortest paths going from node $s$ to node $t$ through node $i$, and $\sigma_{st}$ is the number of all of the shortest paths going from node $s$ to node $t$ (Freeman, 1977). Normalized closeness centrality is defined as $C^C(i) = \frac{N-1}{\sum_{i \neq j} d_{ij}}$, where $d_{ij}$ is the length of the shortest path between node $i$ and $j$ (Freeman, 1979). We explored how betweenness and closeness centralities change as the node strength $str(i)$ increases. Node strengths in each L1Z are scaled to the range of 0 to 1 for comparison. Results show that the 15 L1Zs share common characteristics: betweenness centrality increases linearly with strength as $C^B(i) = k_B str(i) + b_B$ with significant levels of 0.01, and closeness centrality increases linearly with logarithm strength as $C^C(i) = k_C \log(str(i)) + b_C$ with significant levels of 0.001. The increases illustrate that the volume centers are also structure centers, which play decisive roles in organizing local taxi flows. Although residents take taxis directly from origins to destinations instead of transferring in the centers, the increase of betweenness centrality still demonstrates that many low-degree nodes do not have direct interactions between them, while high-strength nodes connect directly to large proportions of nodes in different parts of the L1Z. The increase of closeness centrality shows that high-strength nodes also have closer distances to all of the other nodes in the same L1Z.

In addition to the shared characteristics, some differences exist between urban and suburban L1Zs. Apart from the higher internal trip volumes, the urban L1Zs also have higher graph density values, illustrating that the well-developed urban regions have broader interactions than suburban regions. Moreover, the CCDF curves of suburban L1Zs decrease more quickly in low-strength regions and more slowly in high-strength regions than urban L1Zs, indicating that centers in suburban areas account for larger proportions of internal traffic volumes, which is confirmed by the higher average proportion of internal trips taken by the suburban centers. Additionally, the coefficients $k_B$ and $k_C$ are both larger on average in suburban L1Zs (Fig. 9 (e), (f)), implying that betweenness and closeness centralities grow more quickly in suburban areas. Therefore, low strength nodes have more interactions between each other in urban areas, and centers are more influential on local traffic in suburban L1Zs. The differences may result from the different development levels: the well-developed urban area concentrates various types of resources, generating travels for different purposes among broader regions; land use of suburban regions may be more homogenous, and local travels tend to be dominated by the local centers that contain key resources for this region.

By analyzing sub-networks corresponding to L2Zs, we found that they also have center cells containing large traffic volumes. Despite the fact that L2Zs are merged by L1Zs, only some L1Z centers, which may have higher capacity and also dominate the long-length internal trips, are still acting as centers in L2Zs. Some new centers, in which special functions are particularly attractive to long-distance trips, also emerged in L2Zs. Detailed studies of sub-regions depict a hierarchical polycentric urban structure of Shanghai from the movement flow aspect. The centers stemming from taxi flows show their centrality mostly in the transport system. In addition to centers of market or employment, these centers may also play important transport roles as transfer centers.

Land use indicates the demand of intra-urban travels. To interpret the formation of the polycentric city structure and the way centers interact with local travels, we explored the land use, or the functions in the city, of these centers. Most urban L1Z centers are important commercial and business centers, while suburban L1Z centers favor mixed residential areas and metro stations that are vital for local people. L2Z centers consist of both some of the L1Z centers and important transport terminals, such as airports and train stations. Therefore, the land use of centers illustrate the formation of such hierarchical city structure, which is somewhat consistent with the mechanism mentioned in the Central Place Theory (Christaller, 1966), although at a smaller scale. L1Z centers concentrate important resources for local residents and the type of resources are different for urban and suburban L1Zs. These resources mainly serve the L1Z areas and lead to concentration of local travels. Some of the L2Z centers are L1Z centers that contain the most influential commercial and business resources in the city, serving large L2Z areas. Their influences in intra-city travel decays slower than other L1Z centers. Different from the Central

Place Theory, some irreplaceable transportation facilities also act as L2Z centers, because of their broad attractiveness. The nesting structure indicates the different travel frequency and travel purposes between long and short trips. By revealing the functions of these centers, we have a better understanding of the polycentric urban structure formation and the mechanism for these centers to influence travel patterns in sub-regions. For detailed information about land use of centers and their interaction with sub-regions, please refer to the appendix of the paper.

# 4. Discussion and conclusions

This study uses collective intra-city trips extracted from emerging taxi GPS trajectory data to explore travel patterns and city structure. Network science methods, especially community detection methods, are introduced to reveal the sub-regional city structure on the basis of spatial interactions. This approach reveals the two-level hierarchical polycentric structure of Shanghai and also sheds light on urban and transportation policy-making issues. Some limitations of taxi data representativeness should be noted. First, taxi trips are only able to represent a part of intra-urban travels. Some travels, such as long-distance commuting, are hardly reflected by taxi trajectories. However, most existing data are only able to show city characteristics from a specific perspective. Future studies could take advantage of each type of data and combine the results to show comprehensive patterns of urban spatial interaction. In addition, the conclusions drawn from taxi trip data could also contribute to the improvement of other transportation modes because of their complementary relations. Second, taxi passengers are biased samples of city population. The impact of this issue is mitigated in this study, given that we pay more attention to the properties of L1Zs resulting from local taxi travels which are affordable for, and often preferred by, a majority of dwellers in their daily lives. Therefore, we suggest that taxi trips constitute a relatively stable proportion of intra-city travels with large data volume and high precision. Taxi data are reasonable to represent intra-city spatial interactions and reveal city structure.

This study reveals a two-level hierarchical polycentric structure of Shanghai with a view of spatial interactions represented by taxi trips. Using network science methods to analyze spatial interactions brings a promising approach to the study of city structure from the aspect of traffic flows. The sub-regional structure shows how the places within a city interact and cluster. The sub-regions, viz., the L1Zs and L2Z, could provide new opportunities to determine urban and transportation planning boundaries and help validate existing urban management policies. Most sub-region borders found in this study differed from Shanghai’s city district borders because administrative boundaries have limited impacts on intra-city travels, especially in cities with high dynamics such as Shanghai. Thus, the borders found in this study provide reasonable alternative boundaries for urban planning because the places in the same sub-regions have more interactions than in administrative regions. Traditional transportation planning boundaries, such as TAZs and census blocks, are based on population. The sub-regions we found are based on connectivity strength between small analysis units such as TAZs. The new boundaries could not only provide insights into improving local mobility but also help eliminate unneeded TAZs to improve computational efficiency when modeling transportation. For built facilities or functional regions, the method could be used to validate whether they are serving residents as planned by checking the sub-regions they belong to.

The hierarchical polycentric structure and the explanations of its underlying mechanisms from the land use aspect could also contribute to transportation planning. The sub-regional borders identify the most influential areas of local centers, and policies could be made to improve accessibility to local centers or reduce the total amount of travel. Taxis are an important supplement for other means of public transportation in Shanghai. Given the fixed stops of buses and metros, taxi flows can highlight the areas where traffic demand exceeds the current service levels. Bus route modifications and metro line extensions, especially to local centers, might meet the traffic demands and improve accessibility. Because land use has a strong relation with the travel behaviors of residents, travel amount could also be reduced by modifying land use of L1Zs and their centers. For example, L1Z 11 covers a hi-tech industrial park, and the center contains an important transfer metro station. Large amounts of local flows are generated by people who live in places far from the industrial part and come to the station by taxi after work for cheaper long-distance travels. Planning new residential areas in L1Z 11 could help reduce traffic flows resulting from work–home separation. Moreover, suburban centers in residential areas that contain transfer facilities such as metro stations may also have the potential to be developed into sub-commercial centers. Thus, some long-distance trips could be turned into short local trips, thereby reducing the total travel distance.

The urbanization process in China is rapid, and urban structures of world-level cities such as Shanghai are becoming rather complex. Big data provide us with opportunities to conduct empirical research on residents' mobility patterns and the corresponding urban structures, breaking ground for new theoretical studies and contributing to urban and transportation planning. The methods provided in this study are also suitable for analysis of other cities with similar data sources. Further studies may expand the data source to include private car, bus, and metro trips. This combination of diverse data could describe human mobility and urban structure in more dimensions, providing a multi-faceted picture of urban dynamics. The data source may also be expanded on the time dimension. The availability of long-period human mobility data would make it possible to detect the change of the urban structure and validate the effect of policies.

# Acknowledgements

This research was supported by the Natural Science Foundation of China (Grant nos. 41271386, 41371169, and 41428102). Xi Liu would like to thank Xize Wang for his helpful suggestions.

# Appendix

To interpret the factors influencing intra-city travel patterns and the formation of the polycentric city structure, the land use, or the function in the city system, of centers should be explored. The land uses of most centers are hard to directly identify because land uses are mixed and the center functions are a combination of many factors. Given that the temporal variations of taxi pick-up and drop-off points strongly correlate with the area's function (Liu et al., 2012b), we tried to infer land uses of L1Z centers with temporal pick-up and drop-off volumes. We computed the total number of pick-up and drop-off points of each center for each hour of the day and plotted the results in Fig. A2. Detailed pick-up and drop-off point locations and empirical knowledge of the area were also taken into consideration. For simplicity, in this section, the center of a L1Z is still defined as the node with the highest strength value. This simplification is reasonable because most double/triple centers of the same L1Z are adjacent or have similar functions. The land uses of emerging L2Z centers are easy to identify because they represent important transportation terminals. The centers are depicted in Fig. A1, and the most traffic-influential facilities in each center are listed in Table A1.

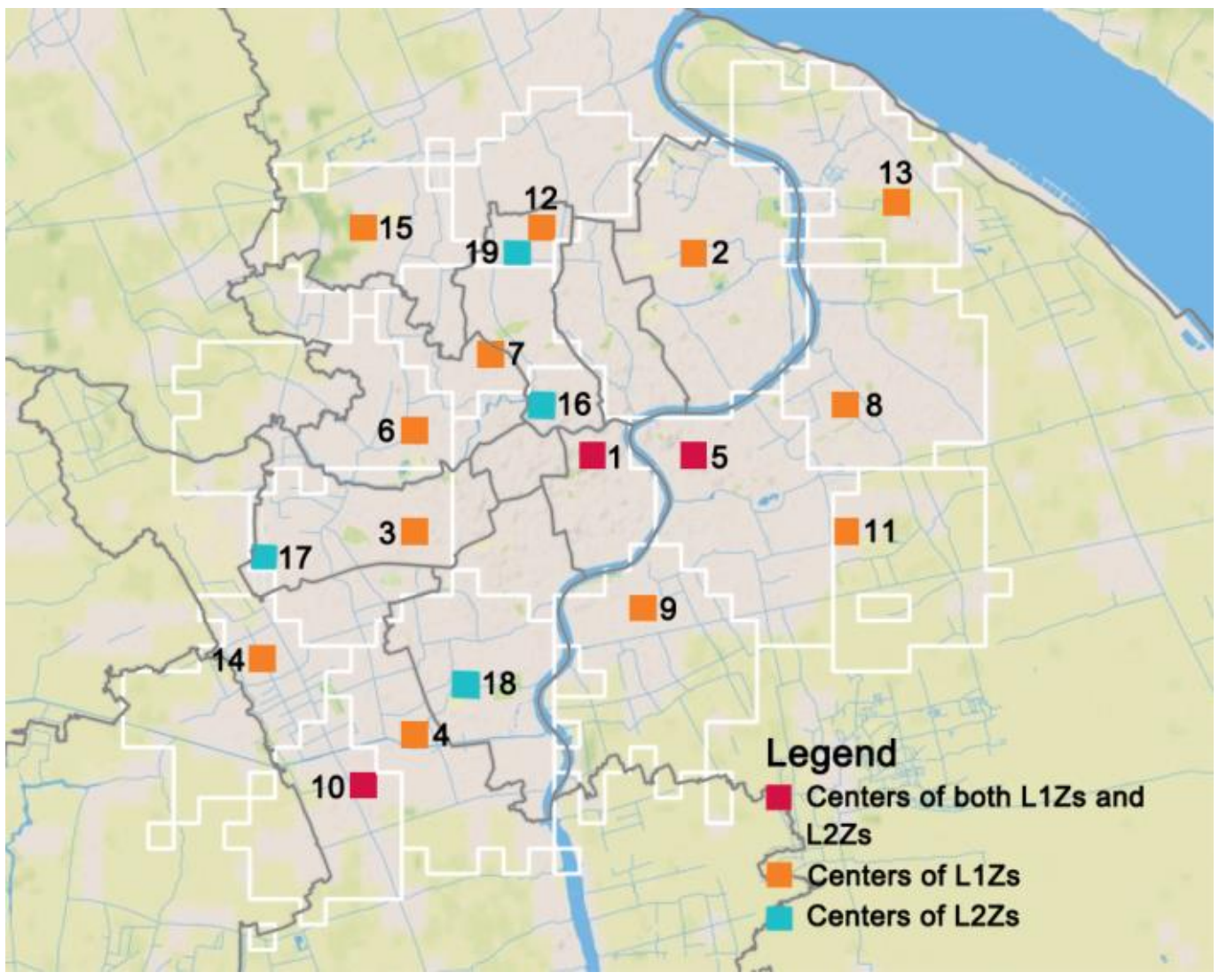

Fig. A1. The centers of selected L1Zs and L2Zs.

Table A1. Land uses of centers

| Centers | Corresponding sub-regions | Most traffic-influential facilities | Land use |
|---|---|---|---|
| 1 | L1Z 1 and L2Z I | Government Building; Nanjing Road; People's Square | Business and commercial area |
| 2 | L1Z 2 | Wujiaochang | Commercial and business area |
| 3 | L1Z 3 | New Century Plaza; Shanghai Mart | Business and commercial area |
| 4 | L1Z 4 | Nanfang Friendship Shopping Center; | Commercial and business |
| | | Lianhua RD Metro Station | area |
| 5 | L1Z 5 and L2Z II | New Shanghai Commercial Area | Commercial and business area |
| 6 | L1Z 6 | Caoyang Shopping Mall; The Central Hospital of Putuo District | Commercial area |
| 7 | L1Z 7 | Residential Areas; Tongji Hospital | Mixed residential area |
| 8 | L1Z 8 | Residential Areas; Malls; Schools | Mixed residential area |
| 9 | L1Z 9 | Yaohua Rd Metro Station | Transportation |
| 10 | L1Z 10 and L2Z III | Xinzhuang Metro Station | Transportation |
| 11 | L1Z 11 | Zhangjiang Hi-tech Park Metro Station | Transportation |
| 12 | L1Z 12 | Residential Areas; Malls; Schools | Mixed residential area |
| 13 | L1Z 13 | South Waigaoqiao Free Trade Zone Metro Station | Transportation |
| 14 | L1Z 14 | Qibao Mall; Residential Areas | Mixed residential area |
| 15 | L1Z 15 | Residential Areas; Malls | Mixed residential area |
| 16 | L2Z I | Shanghai Railway Station | Transportation |
| 17 | L2Z I | Shanghai Hongqiao International Airport | Transportation |
| 18 | L2Z III | Shanghai South Railway Station | Transportation |
| 19 | L2Z IV | Pengpu Xincun Rd Metro Station | Transportation |

We first analyzed centers located in urban L1Zs. The center of L1Z 1 lies in the central district of Shanghai containing the city government, famous commercial areas, and important office buildings, and thus, high levels of taxi flows result during the daytime. The number of pick-up points overwhelms the drop-off points in the evening, demonstrating the trend of people leaving the commercial and business areas. Major functions of L1Z centers 2 to 6 are also commercial and business areas, leading to more people traveling to these regions in the morning and leaving them in the evening. The curves may have different detailed trends and peaks because the ratios between business and commercial areas vary. Some centers also have hospitals or metro stations with high volumes whose influences are also contained in the curves. The center of L1Z 7 is dominated by residential land use mixed with small commercial regions. Because residential areas generate traffic flows in the morning and attract traffic flows in the evening while commercial areas influence traffic in an almost opposite way, people come in and out of this region throughout the day with almost equivalent numbers, leading to two relatively similar curves. The peak of drop-off points in the morning mainly results from the presence of a hospital.

Unlike most centers in urban areas, which act as local commercial and business centers, suburban centers are strongly influenced by residential land use and metro stations. Centers of L1Zs 8, 12, 14, and 15 are residential areas mixed with some commercial areas and public places, such as schools. Similar to the center of L1Z 7, the pick-up point and drop-off point curves closely match each other. Centers of L1Zs 9 and 10 contain metro stations surrounded by residential areas, resulting in peak drop-off points in the morning as residents transfer to metro—the cheaper mode for long-distance travels—and then travel to central urban areas or other places. Conversely, pick-up points continue to grow throughout the daytime and peak in the evening as residents take taxis to leave the terminal and go home. In the center of L1Z 10, the number of people arriving throughout the day is much higher than the number of people leaving because the metro station in L1Z 10 belongs to an integrated public transport terminal. Many people arrive by taxi and transfer to continue their travels. L1Zs 11 and 13 comprise a hi-tech industrial area and a free trade zone, with centers of those two L1Zs both containing important metro stations. People utilize these areas for work and business in the daytime, many of whom tend to take the subway to the stations in these areas and then transfer to taxis to reach their destinations; thus, the pick-up point curve peaks in the morning and then begins to decrease. After work or business, people take taxis to the metro stations and leave the areas, leading to the peaks of drop-off point curves in the afternoon. The higher drop-off point numbers in L1Z 11 may result from the combination of research buildings and commercial centers near the metro station. Furthermore, the higher pick-up numbers in L1Z 13 may result from the limited time budgets of people before business, resulting in the preference to travel by taxi to save time.

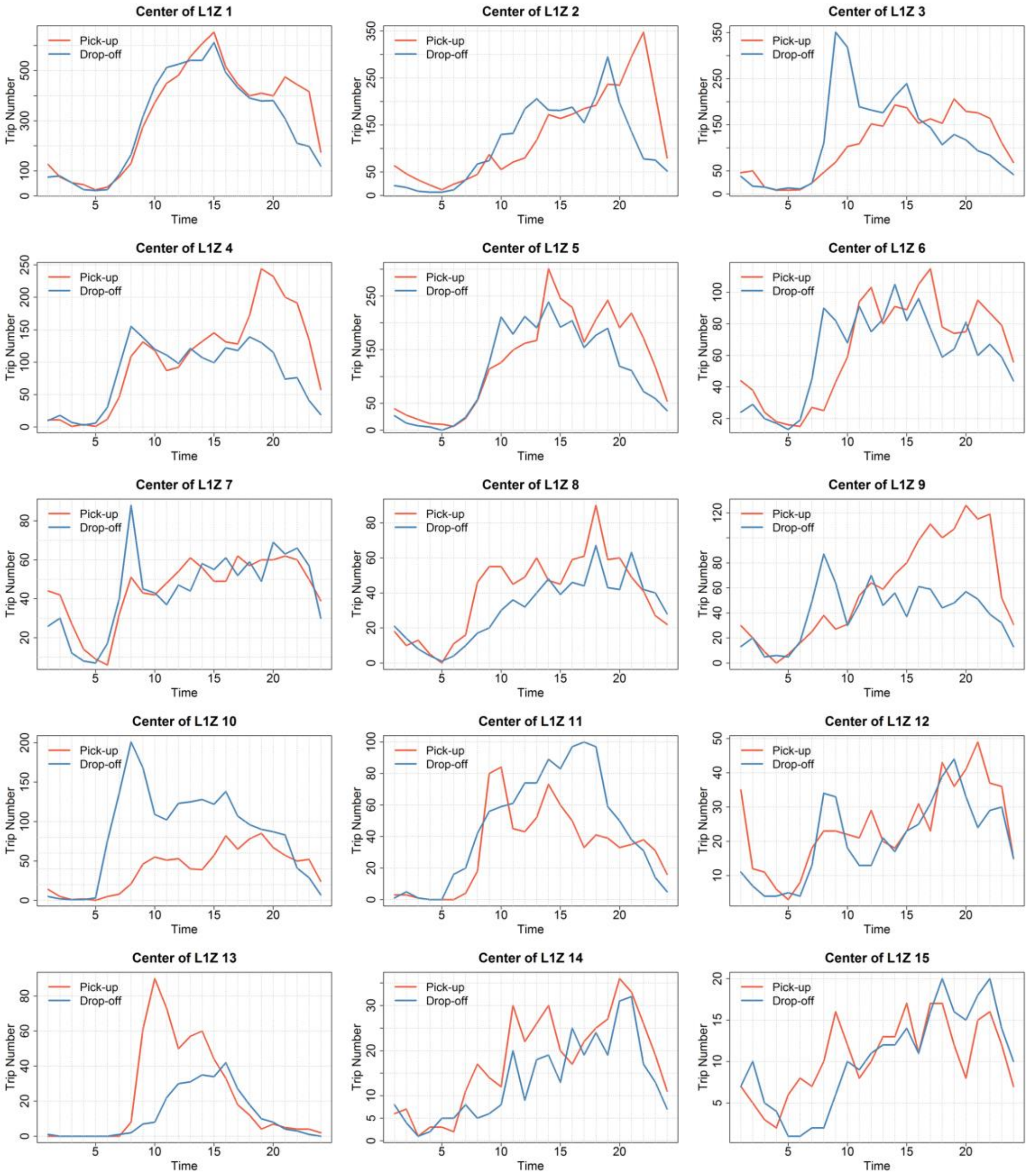

Fig. A2. Temporal variations of pick-up and drop-off points of centers in each L1Z.

Being the center of the entire city, center of L1Z 1 also enjoys intense interaction with cells in nearby L1Zs and continues acting as the center of L2Z I. In addition, the Shanghai Railway Station and Hongqiao International Airport emerge as new centers, given their irreplaceable roles in Shanghai’s transportation system and their strong abilities in attracting and generating long-distance trips. The center of L1Z 5 contains the most influential commercial and business zone in the eastern part of Shanghai, thus being the center of L2Z II. Given that L2Z III primarily covers suburban areas, the center of L1Z 10, with the integrated transport hub, still has broad influence and acts as a center. Shanghai South Railway Station also emerges as a center in L2Z III. The original L1Z centers still act as centers in L2Z IV. Because L1Zs in L2Z IV are primarily residential areas, the cell containing a crucial metro station also emerges as a new center.